\documentclass[12pt,english,aps,manuscript]{article}
\usepackage[T1]{fontenc}
\usepackage[latin1]{inputenc}
\usepackage{geometry}
\usepackage{amsmath}
\usepackage{amssymb}
\usepackage[numbers]{natbib}

\makeatletter

\providecommand{\tabularnewline}{\\}

\usepackage{geometry}

\makeatletter
\usepackage{color}

\makeatletter
\newcommand{\bee}{\begin{equation}}
\newcommand{\ee}{\end{equation}}
\newcommand{\beea}{\begin{eqnarray}}
\newcommand{\eea}{\end{eqnarray}}

\makeatother

\makeatother

\makeatother

\usepackage{babel}
\begin{document}
\begin{center}
\textbf{\Large Gauge Threshold Corrections and Field Redefinitions}{\Large{} }
\par\end{center}{\Large \par}

\begin{center}
\vspace{0.3cm}
 
\par\end{center}

\begin{center}
{\large S. P. de Alwis$^{\dagger}$ } 
\par\end{center}

\begin{center}
Abdus Salam International Centre for Theoretical Physics, Strada Costiera
11, \\
 Trieste 34014, Italy 
\par\end{center}

\begin{center}
and
\par\end{center}

\begin{center}
Physics Department, University of Colorado, \\
 Boulder, CO 80309 USA 
\par\end{center}

\begin{center}
\vspace{0.3cm}

\par\end{center}

\begin{center}
 
\par\end{center}

\begin{center}
\textbf{Abstract} 
\par\end{center}

We review the argument for field redefinitions arising from threshold
corrections to heterotic string gauge couplings, and the relation
between the linear and the chiral multiplet. In the type IIB case
we argue that the necessity for moduli mixing at one loop order has
not been clearly established, since this is based on extending the
background field expansion way beyond its regime of validity. We also
resolve some issues related to the form of non-perturbative terms
resulting from gaugino condensation. This  enables us to estimate
the effective cutoff in the field theory by evaluating the non-perturbative
superpotential by two different methods, and find that it is around
the Kaluza-Klein scale, as one might have expected on general grounds
of self-consistency.

\begin{center}
\vspace{0.3cm}
 
\par\end{center}

\vfill{}

$^{\dagger}$ dealwiss@colorado.edu

\eject

\section{Introduction}

Background field methods in string theory (see for example \citep{Green:1987sp}\citep{Green:1987sp,Polchinski:1998rq,Polchinski:1998rr}
are different from what one encounters in say, the quantum mechanical
study of the behavior of an atom in an external magnetic field. In
the latter case the background magnetic field is truly external to
the system under study. In field theory applied say to condensed matter
physics or atomic physics, where one is studying not the theory of
the entire universe but some system within it, the concept of an external
background makes sense. However when one studies theories such as
the standard model coupled to gravity, which purports to be an effective
theory of the entire universe, strictly speaking there is no meaning
to the concept of an external background.

Of course the standard model is usually formulated in a particular
metric background - namely the flat one. Here the reasoning is that
for small standard model field energy densities, the Einstein equations
are solved by the Minkowski metric. In principle it can be studied
in a different metric background for example a cosmological background
(FRW, deSitter etc.). However this smooth gravitational background
field assumption is certainly expected to break down close to the
Planck scale ($M_{P}\equiv1/\sqrt{8\pi G_{Newton}}$). At such high
energies one expects a significant contribution from virtual quantum
gravity processes (such as the creation and annihilation of blackholes,
wormholes etc.) and the entire framework will break down.

String theory on the other hand is supposed to be an UV completion
of field theory (or at least of a class of field theories hopefully
including the standard model). The theory is not supposed to have
any free parameters and is defined purely in terms of a fundamental
dimensional constant - the string scale $l_{string}\equiv\sqrt{2\pi\alpha'}=1/M_{string}$.
If string theory were four dimensional, $M_{string}$ would essentially
be the same as $M_{P}$. However in all tractable string theoretic
constructions, there is an internal six-dimensional space with a volume
which is typically large compared to $l_{string}^{6}$, i.e. ${\rm Vol}={\cal V}l_{string}^{6}$
often with ${\cal V}\gg1$. In this case there is a significant difference
between the two scales and $M_{S}\simeq M_{P}/\sqrt{{\cal V}}\ll M_{P}$.
There is also an additional scale, the Kaluza-Klein scale $M_{KK}=M_{string}/{\cal V}^{1/6}=M_{P}/{\cal V}^{2/3}$.
For large ${\cal V}$ we thus have a hierarchy of scales $M_{KK}\ll M_{string}\ll M_{P}.$
Four dimensional field theory is strictly valid only below $M_{KK}$.
Above this scale the theory is essentially ten dimensional but remains
a field theory. However above the scale $M_{string}$, the theory
\textit{cannot }be described by point like field theoretic degrees
of freedom. The field theoretic description necessarily breaks down.

Consider first the case of strings propagating in a general metric
background. In this case the world sheet theory is formulated as a
generalized two dimensional sigma model, and one derives consistency
conditions (beta function equations) for the propagation of strings,
in an expansion in the squared string length scale - the so-called
$\alpha'$ expansion. For energy scales that are well below the string
scale, this is a valid expansion and one can get useful information
about the low energy limit of string theory in this way. However this
expansion obviously breaks down at the string scale. In fact this
is highlighted by the fact that the derivative expansion (as with
generic higher derivative theories) has ghosts. These however appear
at the string scale and are merely a sign that the theory has been
pushed beyond its regime of validity. Thus any argument that is made
about the interaction vertices of the theory - that is derived from
the $\alpha'$ expansion - is invalid when the momentum flowing through
those vertices is greater than the string scale. At such energies
the low energy point field theory needs to be replaced by string (field?)
theory.

The same is true for open string background calculations. Here one
turns on a gauge field strength (magnetic or electric) background
that is slowly varying (if not constant), to derive a low energy effective
action in an $\alpha'$ expansion. Much important work has been done
by using this technique. However for the most part this work has been
used only to get an effective field theory valid below the string
scale. In particular it does not make sense to consider the behavior
of a term like $\frac{1}{g^{2}(\mu^{2})}{\rm tr}(F_{u\nu}F^{u\nu})$
when the momentum flowing through this operator is greater than the
string scale. The representation in terms of such a local operator
simply breaks down at these energies. In particular this means that
any argument which purports to have a low (around say few TeV) string
scale and gauge coupling unification at some much higher scale (such
as the standard GUT scale of $10^{16}GeV$), cannot possibly make
sense. 

In this work we will address the question of string threshold corrections
to gauge couplings by first reviewing the literature. The issue came
up with Kaplunovsky's calculation of these effects in the context
of the heterotic string \citep{Kaplunovsky:1987rp}. The question
that arose was how to account for moduli dependent corrections that
could not be written as harmonic functions of the moduli as would
be required in the usual chiral field formulation of the effective
supergravity coming from string theory. The resolution in terms of
the linear and chiral multiplet duality was discussed in \citep{Derendinger:1991hq}\citep{Binetruy:2000zx}
(BGG) and is reviewed in section two, where we also point out that
the argument fails when the modulus in question has a non-linear superpotential
term in the relevant chiral superfield. In the next section we review
the arguments of Kaplunovsky and Louis which compared their field
theoretic formula for the gauge coupling function with the corresponding
string theory calculation. It should be stressed that this only involved
momentum scales below the string scale (as we will make clear below).
By contrast the work of references \citep{Conlon:2009xf,Conlon:2009kt,Conlon:2010ji},
as well as the earlier work of \citep{Bachas:1998kr,Antoniadis:1998ax,Antoniadis:1999ge}
in connection with low scale strings, is essentially based on using
the effective field theory and background string theory approaches
well above the string scale. The point is that in identifying the
infra-red region of the string theory integrand the upper (UV) cut
off is taken beyond the string scale. This is essentially the meaning
of taking the infra-red region of the one-loop string integral all
the way up to and beyond the string scale. The argument for the necessity
of certain field redefinitions that result from this comparison are
then very sensitive to exactly where the cutoffs are located, and
can be changed by appropriately choosing the cutoffs at scales below
the string scale. Finally we resolve a long standing puzzle regarding
two different derivations of the non-perturbative (gaugino condensate)
term in the superpotential.

\section{General framework}

\subsection{Linear -chiral duality $ $with $\partial_{S}W=0$.}

As mentioned earlier the linear multiplet - chiral multiple duality
has been discussed for example in \citep{Derendinger:1991hq} and
\citep{Binetruy:2000zx}(BGG). We essentially follow the discussion
of BGG except that the Kaehler supergravity framework is replaced
by the standard (minimal) one. We begin with the following action
(with $\kappa=M_{P}^{-1}=1,\, d^{8}z=d^{4}xd^{4}\theta,\, d^{6}z=d^{4}xd^{2}\theta$)
for chiral superfields $\Phi$ (having superpotential $W$ and Kaehler
potential $K$) coupled to supergravity and gauge fields with prepotential
$V$ and gauge field strength ${\cal W}$. 
\begin{eqnarray}
{\cal A} & = & -3\int d^{8}z{\bf E}\exp[-\frac{1}{3}K(\Phi,\bar{\Phi};V)]\nonumber \\
 &  & +\left(\int d^{8}z\frac{{\bf E}}{2R}[W(\Phi)+\frac{1}{4}f(\Phi){\cal W}{\cal W}]+h.c.\right).\label{eq:action}
\end{eqnarray}
Here ${\bf E}$ is the full superspace superdeterminant and $R$ is
the chiral superspace curvature. Note that the gauge coupling function
$f$ is in general a \textit{holomorphic }gauge invariant function
of the chiral superfields $\Phi$. However threshold effects in string
theory appeared to give non-holomorphic moduli dependent corrections
to the gauge coupling function. The resolution lay in the introduction
of the linear multiplet formulation of the gauge coupling function.
The key observation here is that string theory moduli and the dilaton
naturally arise in the string theory context as (components of) linear
multiplets. This is because axionic partners of the scalar moduli
are in fact second rank tensor fields. Thus for instance the axio-dilaton
$S$, which is often identified in 4D as a chiral scalar, has its
origins in a multiplet containing an antisymmetric second rank tensor
$b_{\mu\nu}$, and thus naturally belongs to a linear multiplet. 

Let us first focus on this case where in the usual formulation the
gauge coupling function is given by $f=kS$ where $S$ is the dilaton
chiral superfield i.e. $\bar{\nabla}^{\dot{\alpha}}S=0.$ Let $\{\phi\}$
denote all the other chiral superfileds in the theory. We take (for
the moment) the superpotential $W$ to be independent of $S$ as is
the case in perturbative string theory (except in IIB where it can
be linear in $S$ in the presence of internal fluxes). Let $U$ be
an unconstrained real superfield and modify the above action to the
following form (for simplicity we ignore chiral fields which are charged
under the gauge group),
\begin{eqnarray}
{\cal A} & = & -3\int d^{8}z{\bf E}\exp[-\frac{1}{3}K(\phi,\bar{\phi},U)](F(\phi,\bar{\phi},U)+U(S+\bar{S}))\nonumber \\
 &  & +\left(\int d^{8}z\frac{{\bf E}}{2R}[W(\phi)+\frac{1}{4}kS{\cal W}{\cal W}]+h.c.\right).\label{eq:Slinchiral}
\end{eqnarray}
Here a trace over the gauge group is implicit in the gauge kinetic
term, $F$ is a real function of the chiral fields $\phi$ and the
real field $U$, which will be determined by a normalization condition
below. $U$ is introduced as the linear superspace dual of the chiral
superfield $S$ (for more details see \citep{Derendinger:1991hq}
\citep{Binetruy:2000zx}) and as we will see this field becomes useful
in interpreting threshold corrections in string theory. Note that
we have also modified the Kähler potential to include dependence on
$U$. Now we may eliminate the chiral superfield $S$ in favor of
the real superfield $U$, by using the equation of motion coming from
taking the $\delta S$ variation of this action to get; %
\footnote{In taking a variation w.r.t. a chiral field we need to set $\delta S=(-\frac{1}{4}\bar{\nabla}^{2}+2R)\delta\Sigma$
where $\Sigma$ is an unconstrained superfield.%
}
\begin{equation}
-3(-\frac{1}{4}\bar{\nabla}^{2}+2R)(Ue^{-K/3})+\frac{k}{4}{\cal W}^{2}=0.\label{eq:deltaS}
\end{equation}
Note that this equation and its conjugate are now effectively constraints
on the initially unconstrained superfield $U$. In the absence of
the gauge field kinetic term and $K$ this is essentially the linear
superfield constraint. Here we have a modified linear superfield.

Substituting \eqref{eq:deltaS} into \eqref{eq:Slinchiral} we get
the (modified) linear multiplet formulation of the above action
\begin{eqnarray}
{\cal A}_{LMF} & = & -3\int d^{8}z{\bf E}e^{-K(\phi,\bar{\phi},U)/3}F(\phi,\bar{\phi},U)\label{eq:ALMF}\\
 &  & +(\int d^{8}z\frac{{\bf E}}{2R}W(\phi)+h.c.).\nonumber 
\end{eqnarray}
Note that there is no explicit gauge kinetic term in this form of
the action. It is however implicit because $U$ satisfies the constraint
\eqref{eq:deltaS}. To see this we first rewrite this as an equation
for $-\frac{1}{4}\bar{\nabla}^{2}U$ to get (keeping only terms which
contain ${\cal W}^{2}$)
\begin{equation}
-\frac{1}{4}\bar{\nabla}^{2}U=\frac{e^{K/3}}{1-\frac{1}{3}UK_{U}}\frac{k}{3\times4}{\cal W}^{2}+\ldots.\label{eq:del2U}
\end{equation}
Then we have from the first line of \eqref{eq:ALMF}, using $\int d^{8}z{\bf E}\bar{\nabla}^{\dot{\alpha}}v_{\dot{\alpha}}=0$
and the above result, 
\begin{eqnarray*}
-3\int d^{8}z\frac{{\bf E}}{2R}(-\frac{1}{4}\bar{\nabla}^{2}+2R)\left(e^{-K/3}F(\phi,\bar{\phi},U)\right) & =\\
-3\int d^{8}z\frac{{\bf E}}{2R}e^{-K/3}(-\frac{K_{U}}{3}F+F_{U})(-\frac{1}{4}\bar{\nabla}^{2}U)+\ldots & = & -\int d^{8}z\frac{{\bf E}}{2R}\frac{k}{4}\Gamma(U,\phi,\bar{\phi}){\cal W}{\cal W}+\ldots,
\end{eqnarray*}
so that the effective gauge coupling function in the linear multiplet
formulation is (the lowest component of) 
\begin{equation}
\Gamma=-F_{U}+\frac{N}{3}K_{U},\, N\equiv\frac{F-UF_{U}}{1-\frac{1}{3}UK_{U}}\label{eq:Gamma}
\end{equation}
 in agreement with BGG.

Now let us see what we would get in the chiral field formulation of
the gauge coupling function. Varying the action \eqref{eq:Slinchiral}
with respect to $U$ we get
\begin{equation}
(S+\bar{S})(1-\frac{1}{3}UK_{U})=\frac{1}{3}FK_{U}-F_{U}.\label{eq:Ueqn}
\end{equation}
This equation determines $U=U(S+\bar{S},\phi,\bar{\phi})$. In the
chiral field formulation we need to have in addition the normalization
condition
\begin{equation}
F(\phi,\bar{\phi},U)+U(S+\bar{S})=1.\label{eq:norm}
\end{equation}
This is just the condition that in the chiral formulation the pre-factor
of $e^{-K/3}$ should be unity. Substituting into \eqref{eq:Slinchiral}
gives the chiral field formulation with the standard chiral gauge
coupling function i.e. \eqref{eq:action}. Note also that eliminating
$(S+\bar{S})$ between the last two displayed equations gives the
relation 
\begin{equation}
1-\frac{1}{3}UK_{U}=F-UF_{U}\label{eq:norm2}
\end{equation}
 so that the function $N=1$ (see \eqref{eq:Gamma}) and the expression
for $\Gamma$ becomes 
\begin{equation}
\Gamma=-F_{U}+\frac{1}{3}K_{U}.\label{eq:Gamma2}
\end{equation}

Equation \eqref{eq:norm2} can be rewritten as a differential equation
for F,
\[
-U^{2}\frac{d}{dU}(U^{-1}F)=1-\frac{1}{3}UK_{U}.
\]
This has the solution
\begin{equation}
F=1+\frac{U}{3}\int^{U}\frac{dU'}{U'}K_{U'}+V(\phi,\bar{\phi})U.\label{eq:Ssoln}
\end{equation}
A simple example that is relevant to string theory is obtained by
putting 
\begin{equation}
K=\hat{K}(\phi,\bar{\phi})+\alpha\ln U.\label{eq:Kexample}
\end{equation}
 In this case $F=1-\frac{\alpha}{3}+VU$ and $\frac{\alpha}{3}U^{-1}=S+\bar{S}+V(\phi,\bar{\phi})$.
Then in the chiral multiplet formulation we have the action \eqref{eq:action}
with $f=kS$ and 
\begin{equation}
K=\hat{K}(\phi,\bar{\phi})-\alpha\ln(S+\bar{S}+V(\phi,\bar{\phi})).\label{eq:KexampleCMF}
\end{equation}
In the string theory case where $S$ is the four dimensional dilaton
chiral superfield $\alpha=1$ and $V$ is a one-loop effect.

On the other hand we have in the linear multiplet formulation the
action \eqref{eq:ALMF} with Kaehler potential \eqref{eq:Kexample},
the gauge coupling (see \eqref{GammaEffective}) given by (putting
$\alpha=1)$ 
\begin{equation}
\Gamma=-\frac{1}{3U}+V(\phi,\bar{\phi}).\label{eq:Gammamodel}
\end{equation}
 The moral of this story is that if one wants to accommodate a non-harmonic
gauge coupling function (as in the above relation) one must choose
the LM formulation. On the other hand this is equivalent to taking
the chiral multiplet dual field $S$ as the gauge coupling function
in the CM formulation but with the non-harmonic part $V$ included
in the Kähler potential for $S$.

\subsection{\label{sub:d_SWne0}$\partial_{S}W\ne0$}

The discussion in the previous subsection remains valid if the superpotential
dependence on $S$ is no more than linear. This is the case for instance
in type IIB string theory, provided the pre-factors of the non-perturbative
terms responsible for breaking the no-scale structure and stabilizing
the Kaehler moduli are independent of $S$ (or at least no more than
linear in $S$). Thus under these usual assumptions one has 
\begin{equation}
W=A(\phi)+B(\phi)S,\label{eq:WIIB}
\end{equation}
where as in the last subsection $\phi$ stands for all the other chiral
scalar superfields. In this case the only relevant change is in equation
(\ref{eq:deltaS}), which is replaced by 
\[
-3(-\frac{1}{4}\bar{\nabla}^{2}+2R)(Ue^{-K/3})+\frac{k}{4}{\cal W}^{2}+B(\phi)=0.
\]
Equation \eqref{eq:ALMF} will remain the same with $W\rightarrow A(\phi)$
but on evaluating the components of the first term we would need to
replace $k{\cal W}^{2}\rightarrow k{\cal W}^{2}+B(\phi)$. The main
results of the previous subsection remain unchanged. 

However when the superpotential is non-linearly dependent on $S$
these results cannot be obtained. The reason is that the $\delta S$
equation no longer gives a constraint on the unconstrained superfield
$U$. Instead it becomes an equation which expresses the chiral superfield
S in terms of a chiral projection of an unconstrained superfield.
Defining ${\bf P}=(-\frac{1}{4}\bar{\nabla}^{2}+2R)$ equation \eqref{eq:deltaS}
is replaced by ($\partial_{S}W\equiv W_{S}$), 
\begin{equation}
-3{\bf P}(Ue^{-K/3})+\frac{k}{4}{\cal W}^{2}+W_{S}=0.\label{eq:WSconstraint}
\end{equation}
Thus we need to replace $S$ in equation \eqref{eq:Slinchiral} by
\begin{equation}
S=W_{S}^{-1}(3{\bf P}(Ue^{-K/3})-\frac{k}{4}{\cal W}^{2}).\label{eq:SUtilde}
\end{equation}
Note that in the above $W_{S}^{-1}$ is the functional inverse of
$W_{S}$.

Substituting \eqref{eq:WSconstraint} into \eqref{eq:Slinchiral}
(with $W(\phi)\rightarrow W(\phi)+\Delta W(S)$ we have for the chiral
superspace terms
\begin{equation}
\int d^{8}z\frac{{\bf E}}{2R}[W(\phi)+(\Delta W(S)-S\partial_{S}\Delta W)]+h.c.\label{eq:ASUtilde}
\end{equation}
 where $S$ is now given as a function of $U$ by \eqref{eq:SUtilde}.
Clearly gauge kinetic terms are hidden in this expression, but the
coupling functions are essentially power series in ${\bf P}(Ue^{-K/3})$
which can always be rewritten as a chiral projection of some redefined
unconstrained real field $\Omega$ i.e. as ${\bf P}\Omega$. For instance
in the simplest case where $\Delta W=S^{2}/2$ the expression \eqref{eq:ASUtilde}
becomes 
\begin{eqnarray*}
\int d^{8}z\frac{{\bf E}}{2R}[W(\phi)-\frac{1}{2}(3{\bf P}(Ue^{-K/3})-\frac{k}{4}{\cal W}^{2})^{2}]+h.c. & \sim\\
\int d^{8}z\frac{{\bf E}}{2R}[W(\phi)-(3{\bf P}(Ue^{-K/3})\frac{k}{4}{\cal W}^{2}] & + & h.c.
\end{eqnarray*}
 so that the gauge coupling function is $3{\bf P}(Ue^{-K/3})$. As
can be easily seen this is the case whatever the functional form $\Delta W$
takes, since this relates to the coefficient of the leading term in
the expansion in ${\cal W}^{2}$. This is obviously of the same form
as the original chiral field representation in terms of $S$ for the
gauge coupling function, and is very different from the linear multiplet
formulation where it was precisely the constraint on $U$ that enabled
us to write a non-holomorphic gauge coupling.

What this illustrates is the general expectation that the dualization
makes sense only when there when the dualized field pair represents
a massless state. The linear superpotential case studied at the beginning
of this subsection is thus an exception.

\subsection{Generalization}

The discussion in the previous two subsections can be easily generalized
to the case when there is a duality relation between several chiral
multiplets $S^{i},\, i=0,\ldots,N$ and linear multiplets $U_{i}$.
For future reference in the type IIB string theory case we rewrite
these fields as $S\equiv S^{0}$ and $T^{i}\equiv S^{i},\, i=1,\ldots,N$.
We also choose to dualize only the $T'$s and write the dual fields
as $U_{i},\, i=1,\ldots,N$. The gauge field kinetic term (in the
chiral formulation) is now taken to be $f_{a}{\cal W}_{a}^{2}$ where
a sum over the different gauge groups is implied, and 
\[
f_{a}=k_{a}S+\alpha_{ai}T^{i}
\]
 One follows the same steps from eqn \eqref{eq:ALMF} through \eqref{eq:norm}
replacing the appropriate products by dot products etc. to get after
dualization of the $T'$s the gauge coupling function,
\begin{equation}
k_{a}S+\Gamma_{a},\,\Gamma_{a}=\Gamma^{i}\alpha_{ai},\,\Gamma^{i}=-F_{U_{i}}+\frac{1}{3}K_{U_{i}}.\label{eq:Gamma3}
\end{equation}
Note that we've set the normalization function 
\begin{equation}
N\equiv\frac{F-U_{i}F_{U_{i}}}{1-\frac{1}{3}U_{i}K_{U_{i}}}=1,\label{eq:N}
\end{equation}
in the above to get as before the standard SUGRA frame. Also the relation
between the dualized variables is now given by 
\begin{equation}
T^{i}+\bar{T}^{i}=\frac{1}{3}K_{U_{i}}-F_{U_{i}}=\Gamma^{i}(U,S,\phi,\bar{\phi})\label{eq:TUrelation}
\end{equation}
Given a model for $K$ as a function of $U^{i}$ the differential
equation \eqref{eq:N} can be solved. For instance in the case of
type IIB string theory on a Calabi-Yau the (scalar components of)
the linear multiplet variables are related to the two cycle volumes
$v_{i}$ by $U_{i}=v_{i}/{\cal V}$ \citep{Grimm:2004uq} and 
\begin{eqnarray}
K & = & -\ln(S+\bar{S})-\hat{K}(z,\bar{z})-2\ln{\cal V},\label{eq:K1}\\
 & = & -\ln(S+\bar{S})-\hat{K}(z,\bar{z})+\ln\hat{{\cal V}}.\label{eq:K2}
\end{eqnarray}
Here ${\cal V}=\frac{1}{6}\int J^{3}=\frac{1}{6}c^{ijk}v_{i}v_{j}v_{k}$,
$\hat{{\cal V}}=1/{\cal V}^{2}=\frac{1}{6}c^{ijk}U_{i}U_{j}U_{k}$,
with the indices running over $1,\ldots,h_{+}^{(1,1)}$ i.e. the Kähler
moduli that survive the orientifold projection, $z$ stands for the
set of complex structure moduli, and $S=e^{-\varphi}+iC_{0}$, where
$\varphi$ is the dilaton and $C_{0}$ is the RR zero-form. We also
have in this case $F=U_{i}V^{i}(\phi,\bar{\phi})$ where $V^{i}$
are a set of arbitrary real functions and%
\footnote{We ignore the additional dependence on moduli coming from the two
form fields.%
} $\phi=\{z,S\}$. It is important for future reference to note that
these arbitrary functions are independent of the Kähler moduli. Using
these we have from \eqref{eq:TUrelation}
\begin{equation}
T^{i}+\bar{T}^{i}=\frac{2}{3}\tau^{i}-V^{i}(z,\bar{z};S,\bar{S})=\Gamma^{i}(U,S,\bar{S};z,\bar{z}),\label{eq:TtauGamma}
\end{equation}
where $\tau^{i}=c^{ijk}v_{j}v_{k}=c^{ijk}U_{j}U_{k}/\hat{{\cal V}}$
is the volume of the $i$th 4-cycle. 

The results of this subsection are summarized in the following table.

\[
\]

~~~~~~~~~~~~~~~~~~%
\begin{tabular}{|c|c|c|}
\hline 
 & chiral & linear\tabularnewline
\hline 
\hline 
$K$ & $-2\ln{\cal V}(T+\bar{T}+V)$ & $\ln\hat{{\cal V}}$\tabularnewline
\hline 
2$\frac{1}{g_{a}^{2}}$ & $k_{a}(S+\bar{S})+\alpha_{ai}(T^{i}+\bar{T}^{i})$ & $k_{a}(S+\bar{S})+\alpha_{ai}(\frac{2}{3}\frac{c^{ijk}U_{j}U_{k}}{\hat{{\cal V}}}-V^{i})$\tabularnewline
\hline 
\end{tabular}

\[
\]

It is important to note that the functions $V^{i}$ do not depend
on the chiral moduli $T$ which have been dualized. Also even though
these functions are arbitrary at the level of these duality transformation,
the usefulness of these results lies in the fact that such functions
may arise in string perturbation theory. In this case of course (at
one loop level) they are expected to be independent of $S$.

\subsection{Relation to KL formula}

What is the relation if any to the KL formula of the above? 

The KL formula \citep{Kaplunovsky:1994fg} for the gauge coupling
constant for a (simple or $U(1)$ gauge group $G_{a}$) is 
\begin{eqnarray}
\frac{1}{g_{a}^{2}(\Phi,\bar{\Phi};\mu^{2})} & = & \Re f_{a}(\Phi)+\frac{b_{a}}{16\pi^{2}}\ln\frac{\Lambda^{2}}{\mu^{2}}+\frac{c_{a}}{16\pi^{2}}K(\Phi,\bar{\Phi})\nonumber \\
 &  & +\frac{T(G_{a})}{8\pi^{2}}\ln\frac{1}{g_{a}^{2}(\Phi,\bar{\Phi};\mu^{2})}-\sum_{r}\frac{T_{a}(r)}{8\pi^{2}}\ln\det Z_{(r)}(\Phi,\bar{\Phi};g^{2}(\mu^{2})).\label{eq:KL}
\end{eqnarray}
Here $T_{a}(r)={\rm Tr}_{(r)}T_{a}^{2},\, T(G_{a})=T_{a}(r={\rm adjoint})$,
\begin{eqnarray}
b_{a} & = & \sum_{r}n_{r}T_{a}(r)-3T(G_{a}),\label{eq:ba}\\
c_{a} & = & \sum_{r}n_{r}T_{a}(r)-T(G_{a}).\label{eq:ca}
\end{eqnarray}
$\Phi$ stands for a set of neutral fields that will be identified
with string theory moduli $(M)$ and axio-dilaton $(S)$. The total
Kaehler potential has been expanded in powers of the charged matter
chiral fields $Q$ which also defines the metric $Z_{(r)}$ of their
kinetic terms in the representation $r$ of the gauge group, i.e.
\begin{equation}
K(\Phi,\bar{\Phi};Q,\bar{Q})=\kappa^{-2}K(\Phi,\bar{\Phi})+\sum_{r}Z_{(r)\bar{I}J}(\Phi,\bar{\Phi})\bar{Q}_{(r)}^{\bar{I}}e^{2V}Q_{(r)}^{J}+\ldots.\label{eq:Kexpn}
\end{equation}
The above formula was derived within an effective supergravity context
on the understanding that at some scale above the effective cutoff
$(\Lambda)$, this field theory would have to be replaced by its ultra-violet
completion. Here the latter will be assumed to be string theory.

Several comments about the formula \eqref{eq:KL} are in order here.
The first term on the RHS is the (real part of the) holomorphic gauge
coupling of the theory defined in \eqref{eq:action}. Its functional
form may be read off from the relevant string theory whose low-energy
effective action is being studied. It will also include the holomorphic
corrections coming from integrating out massive string states. The
second term is the usual field theory running from the cutoff scale
down to the scale $\mu$. The fourth term comes from the rescaling
anomaly that comes when the gauge field prepotential is redefined
so as to get (from the standard superspace form in \eqref{eq:action})
to the canonically normalized form of the gauge/gaugino field kinetic
terms \citep{ArkaniHamed:1997mj}. The fifth term comes from the Konishi
anomaly whose origin is in the the rescaling necessary to get the
canonical kinetic terms for the chiral scalar/fermion fields. These
all occur already in global SUSY and together constitute the (integrated
form of) the NSVZ beta function equation. In particular they have
nothing to do with rescaling the (super) metric and only involve rescaling
the fields $Q$ and $V$. By contrast the third term comes from the
anomaly that occurs when performing the Weyl transformations (on the
supermetric) that are necessary to go to the Einstein-Kaehler frame
starting from the superspace frame of \eqref{eq:action}. 

However we have derived the relation between the linear multiplet
and formulations entirely within the context of the original SUGRA
frame (i.e. where the Einstein term occurs with the factor $\exp(-K/3)$).
This is in contrast to the formulation in Kähler superspace where
the formulation is directly in the Einstein frame. However the latter
formulation has an extended symmetry including a $U(1)_{A}$ which
is quantum mechanically anomalous. This anomaly of course is the source
of the last three terms in \eqref{eq:KL}. In any case the point is
that the linear multiplet chiral multiplet relation discussed in the
previous subsections have nothing to do with these terms, which come
from transforming from the SUGRA frame to the Einstein frame (and
getting to canonically normalized matter and gauge kinetic terms).

\section{String theory considerations}

The relation between string theory calculations, and the KL formula
was discussed at length in \citep{Kaplunovsky:1995jw}. The string
theory formula takes the general form,

\begin{equation}
g_{a}^{-2}(\mu^{2};S,\bar{S},M,\bar{M})=k_{a}g_{string}^{-2}(S,\bar{S},M,\bar{M})+\frac{b_{a}}{16\pi^{2}}\ln\frac{M_{string}^{2}}{\mu^{2}}+\frac{\Delta_{a}(M,\bar{M})}{16\pi^{2}}.\label{eq:g-2string}
\end{equation}
The term multiplying $k_{a}$ here is (the inverse square of) the
gauge coupling at the string scale and is a sum of the classical term
and a universal threshold correction $\Delta^{{\rm univ}}$. $M$
represents all the moduli. The last term is a possible non-universal
threshold correction so that the total threshold correction is of
the form. 
\begin{equation}
\tilde{\Delta}_{a}=\Delta_{a}+k_{a}\Delta^{{\rm univ}}.\label{eq:Deltatilde}
\end{equation}
The main claim of \citep{Kaplunovsky:1995jw} is that the difference
between $\tilde{\Delta}$ and the non-holomorphic terms in the KL
formula, should be a holomorphic one loop correction to the gauge
coupling. Thus the key relation between the two is 
\begin{equation}
\partial_{M}\partial_{\bar{M}}\tilde{\Delta}_{a}=\partial_{M}\partial_{\bar{M}}[c_{a}\hat{K}(M,\bar{M})-\sum_{r}T_{a}(r)\ln\det Z_{(r)}(M,\bar{M})].\label{eq:harmonic}
\end{equation}
While much of the discussion concerns heterotic string theory the
above is based on general arguments that apply to all string theories
(or for that matter to any UV completion of field theory with heavy
states). As pointed out in the introduction to this paper, {}``the
non-harmonicity (of the physical couplings) is a purely low-energy
effect and can be calculated from the low-energy EQFT (effective quantum
field theory) without any knowledge of the superheavy particles; it
is the harmonic terms in the moduli-dependent effective gauge couplings
that are sensitive to the physics at the high- energy threshold. Such
terms can always be interpreted as threshold corrections to the Wilsonian
gauge couplings $f_{a}(\Phi)$,...''.

The intuition behind this statement is that once the string coupling
is identified in terms of low energy fields the non-universal terms
in gauge couplings coming from integrating out high energy fields
should preserve the structure of SUGRA. Kaplunovsky and Louis go on
to check this in explicit heterotic string theory examples. If this
is violated in typeI/IIB theories then one would need to understand
why this intuition is violated.

\subsection{Heterotic case\label{sub:Heterotic-case:-KL} }

The expression \eqref{eq:KL} is expected to be valid to all orders
in perturbation theory (at least in some renormalization scheme) but
we will focus only on the one loop result. The holomorphic gauge coupling
is 
\begin{equation}
f_{a}(S,M)=k_{a}S+\frac{1}{16\pi^{2}}f_{a}^{(1)}(M).\label{eq:f1loop}
\end{equation}
The first term is the classical (universal) gauge coupling and the
second is the holomorphic one-loop correction. Of course $f$ receives
no further corrections. The other terms are already explicitly of
at least one-loop order so that $K,g_{a}^{2},Z$ can be replaced by
their classical values. Thus we put
\begin{eqnarray*}
K(\Phi,\bar{\Phi}) & = & -\ln(S+\bar{S})+\hat{K}(M,\bar{M})+O(\frac{1}{16\pi^{2}\Re S}),\\
Z_{\bar{I}J}(\Phi,\bar{\Phi}) & = & Z_{\bar{I}J}^{(0)}(M,\bar{M})+O(\frac{1}{16\pi^{2}\Re S}),\\
\frac{1}{g^{2}(\mu^{2})} & = & \Re S+O(\frac{1}{16\pi^{2}}).
\end{eqnarray*}
It is important to note that in the heterotic case the classical 4
dimensional string coupling is defined as 
\[
\frac{1}{g_{string}^{2}}=\Re S=e^{-2\phi}{\cal V}
\]
where ${\cal V}$ is the volume of the internal space in string units
and $e^{\phi}$ is the 10 dimensional string coupling. Then we get
from \eqref{eq:KL}
\begin{eqnarray}
\frac{1}{g_{a}^{2}(\Phi,\bar{\Phi};\mu^{2})} & = & k_{a}\Re S+\frac{b_{a}}{16\pi^{2}}(\ln\frac{\Lambda^{2}}{\mu^{2}}-\ln\Re S)\nonumber \\
 &  & +\frac{1}{16\pi^{2}}[\Re f_{a}^{(1)}(M)+c_{a}\hat{K}(M,\bar{M})-\sum_{r}2T_{a}(r)\ln\det Z_{(r)}^{(0)}(M,\bar{M})].\label{eq:g1loopHet}
\end{eqnarray}
This is to be compared with the string 1 loop calculation \eqref{eq:g-2string}
\citep{Kaplunovsky:1995jw} %
\footnote{See also \citep{Nilles:1997vk} where an additional universal contribution
is identified.%
} where 
\begin{equation}
\frac{1}{g_{string}^{2}(\Phi,\bar{\Phi})}=\Re S+\frac{\Delta^{univ}(M,\bar{M})}{16\pi^{2}}\label{eq:gstring}
\end{equation}
Now in the Heterotic string $M_{string}^{2}=M_{P}^{2}/\Re S$ up to
$S,M$ independent constants. So it appears that the $S$ dependence
of the two formulae \eqref{eq:g1loopHet} \eqref{eq:g-2string} agrees.
It should be stressed here that although in \citep{Kaplunovsky:1995jw}
the cut-off $\Lambda$ is identified with $M_{P}$ this is merely
a matter of convenience, and as the authors observed all that is required
is that in the field theory expression the cutoff $\Lambda$ should
be chosen in an $S,M$ independent manner.

As is well known the natural superspace variable that corresponds
to the axio-dilaton of string theory is the linear multiplet (since
the four dimensional axion is originally a two-form field). Suppose
for the moment we ignore the (non-universal) second line \eqref{eq:g1loopHet}
as well as the last term on the RHS in \eqref{eq:g-2string}. As discussed
in the previous section, at least as long as we do not include a superpotential
term for the chiral superfield $S$ in the chiral multiplet form (CMF)
action%
\footnote{Actually as observed in subsection \ref{sub:d_SWne0} one could have
a linear term - but in the heterotic case there is no such term. The
only possible dilaton superpotential comes form gaugino condensation
and is a sum of exponentials in $S$. %
}, the linear multiplet form is equivalent to the chiral multiplet
form. According to the discussion there, the gauge coupling function
(at $\mu^{2}=M_{string}^{2}$) is to be interpreted initially in the
LMF, i.e. the model \eqref{eq:Kexample} \eqref{eq:Gammamodel}. Here
$U$ should correspond to the dilaton in the linear multiplet formulation
which to zero loop order can be identified as $-(3\Re S)^{-1}$. The
term $V$ in \eqref{eq:Gammamodel} is identified with $\frac{\Delta^{univ}(M,\bar{M})}{16\pi^{2}}$.
In the corresponding chiral multiplet formulation (in terms of $S$)
the instruction implied by this is in effect to drop this term from
the expression for the coupling function (thus identifying at the
high scale $g_{a}^{-2}=k_{a}\Re S$) but include it as a one-loop
correction to $K$ (see\eqref{eq:KexampleCMF}) with 
\begin{equation}
K=-\ln(S+\bar{S}+\frac{\Delta^{univ}(M\bar{M})}{16\pi^{2}})+\hat{K}(M,\bar{M}).\label{eq:K1lphet}
\end{equation}

Note that if one makes the above correction to $K$ and then plugs
that back into the CMF form of the action \eqref{eq:action}, then
all that changes in the KL formula \eqref{eq:KL} is the explicit
expression for the term $\frac{c_{a}}{16\pi^{2}}K(\Phi,\bar{\Phi})$
which now becomes $\frac{c_{a}}{16\pi^{2}}$ times the RHS of \eqref{eq:K1lphet}.
However in \eqref{eq:KL} this is already a one-loop effect so this
one-loop change in $K$ is not going to affect the KL formula to one
loop.

\subsection{Type IIB case}

Again we start with the 1-loop form of the KL formula \eqref{eq:KL}
but now we write 
\begin{equation}
K^{(0)}=-2\ln{\cal V}+\tilde{K}((S,\bar{S};z,\bar{z}).\label{eq:K0}
\end{equation}
We have separated the Kaehler moduli dependence (through the internal
volume ${\cal V}$) from the complex structure ($z$) dependence.
Note that here in contrast to the heterotic case the appropriate axio-dilaton
field in the orientifolded type IIB case is 
\[
S=e^{-\phi}+ia
\]
where $e^{\phi}$ is the 10 dimensional string coupling, and $a$
is the RR zero form field (see for example \citep{Polchinski:1998rr}).
As in the LVS models discussed in \citep{Blumenhagen:2009gk}\citep{deAlwis:2009fn}
it is assumed that the MSSM is located on D3 branes at a singularity
(or D7 branes wrapping a collapsing cycle). For these local models
(or in fact for any no-scale like embedding of the matter sector valid
even in the heterotic case) one can write $Z_{\bar{I}J}^{(0)}=\frac{1}{{\cal V}^{2/3}}\delta_{\bar{I}J}$
to leading order in the large volume expansion. 

Then the one-loop coupling function becomes
\begin{eqnarray}
\frac{1}{g_{a}^{2}(\Phi,\bar{\Phi};\mu^{2})} & = & \Re f_{a}(\Phi)+\frac{b_{a}}{16\pi^{2}}\ln\frac{\Lambda^{2}{\cal V}^{-2/3}}{\mu^{2}}+\frac{c_{a}}{16\pi^{2}}\tilde{K}((S,\bar{S};U,\bar{U})\nonumber \\
 &  & +\frac{T(G_{a})}{8\pi^{2}}\ln\Re f_{a}(\Phi).\label{eq:g_aIIB}
\end{eqnarray}
The (holomorphic) gauge coupling functions are given by,
\[
f_{a}=k_{a}S+\alpha_{ai}T^{i}
\]
In the case at hand the coefficient matrix $\alpha_{ai}\ne0$ only
for the index values $i$ corresponding to the the 4-cycles/dell Pezzo
surfaces on which the standard model 7-branes are wrapped. In the
particular examples that are relevant here, with D3 branes at a singularity,
the relevant cycles shrink to a singularity. 

In \citep{Conlon:2009xf} the cutoff $\Lambda$ is taken to be $M_{P}$.
As we discussed before there is no physical significance in the actual
value of this cut off as long as it is taken to be independent of
the moduli. It is simply the upper limit from which one expects the
RG evolution to start, and if the field theory is a low energy effective
theory whose UV completion is string theory then this scale must necessarily
be less than the string scale. As discussed in \citep{Kaplunovsky:1995jw},
for the purpose of comparing the moduli dependence of this formula
to that coming from string theory, it does not matter at what point
$\Lambda$ is chosen (as long as it is above all low energy thresholds)
and indeed without loss of generality it can be chosen to be $M_{P}$.
This obviously does not imply any statement about the validity of
the field theory up to this point and all the conclusion of that paper
would still hold even if the arbitrary cutoff $\Lambda$ is kept.

The conclusion of \citep{Conlon:2009xf} (see also \citep{Conlon:2009kt})
however depends crucially on the choice $\Lambda=M_{P}$. With this
the argument of the log in the second term of the \eqref{eq:g_aIIB}
is 
\begin{equation}
\frac{\Lambda^{2}{\cal V}^{-2/3}}{\mu^{2}}\rightarrow\frac{M_{P}^{2}{\cal V}^{-2/3}}{\mu^{2}}\simeq\frac{M_{string}^{2}R^{2}}{\mu^{2}},\label{eq:conlon}
\end{equation}
where we've used (the approximate) formula $M_{string}^{2}\simeq M_{P}^{2}/{\cal V}$
and put $R\equiv{\cal V}^{1/6}$ the size of the internal space in
string units. From this it is concluded that the effective unification
scale (for large volume compactifications) can be far above the string
scale%
\footnote{Note that to have unification one not only needs $f_{a}\propto k_{a}$
but also that the third and fourth terms on the RHS of \eqref{eq:g_aIIB}
should be negligible.%
}. As the authors themselves point out in a footnote it is not clear
what the ``operational'' meaning of this is. Clearly it cannot have
any, for if instead of the above choice of $\Lambda=M_{P}$ we took
a fixed value of $\Lambda\lesssim M_{string}$ (as one should)%
\footnote{The RHS of this inequality is of course moduli dependent. So what
this requires is that one has to first decide on the values over which
the volume is allowed to range and then fix $\Lambda$ below the lowest
allowed value.%
}, then the corresponding unification scale would be below the string
scale! However in \citep{Conlon:2009kt} the authors go on to derive
physical conclusions based on having a cutoff that is larger than
the string scale. 

These come from the authors' comparison of \eqref{eq:g_aIIB} (after
replacing $\Lambda\rightarrow M_{P}$) with a string theory calculation.
Based on a background expansion of the sort that we discussed in the
introduction (but extending its infra-red region even beyond the string
scale), the following equation is obtained in \citep{Conlon:2009kt},
\begin{equation}
\frac{1}{g_{a}^{2}(\mu^{2})}=\frac{1}{g^{2}}|_{0}+\beta_{a}\ln\frac{M_{string}^{2}}{\mu^{2}}+\beta_{a}^{{\cal N}=2}\ln\frac{M_{X}^{2}}{M_{string}^{2}}.\label{eq:cp}
\end{equation}
Here $\beta_{a}=\beta_{a}^{{\cal N}=1}+\beta_{a}^{{\cal N}=2}=\frac{b_{a}}{16\pi^{2}},$
corresponding to the contributions to the beta function from ${\cal N}=1,2$
states and $M_{X}^{2}=R^{2}M_{string}^{2}$. The last two terms on
the RHS of the above equation are obtained from cutting off two UV
divergent integrals of the form
\begin{equation}
\beta_{a}^{{\cal N}=1}\int^{\mu^{-2}}\frac{dt}{t},\,\beta^{{\cal N}=2}\int^{\mu^{-2}}\frac{dt}{t}.\label{eq:stringdivergence}
\end{equation}
 The upper limit of these integrals is set by the infra-red RG scale
$\mu$ of the previous discussion. On the other hand the UV cut-off
is identified in the first case with $M_{string}^{-2}$ while in the
second case it is identified with the (even smaller!) winding scale
$M_{X}^{-2}$. As we've pointed out in this note, one should not use
the (infra-red regime of the) background field calculation up to (and
beyond) the string scale. Thus this kind of argument cannot be used
to make any statement about the modular dependence, since all that
one needs to have agreement with the field theory calculation is to
choose the UV cutoff in the above integrals to be the same, and to
take the value
\begin{equation}
\Lambda_{string}^{2}=\frac{\Lambda^{2}}{{\cal V}^{2/3}}<\frac{M_{string}^{2}}{{\cal V}^{2/3}}.\label{eq:stringcutoff}
\end{equation}
This cutoff is (for ${\cal V}>1$) obviously well within the regime
of validity of the background field method. In other words there is
really no new information in this calculation. It just gives us the
translation to string language of the choice of cutoff in the field
theory. All that is required for consistency is that both are well
under the string scale!

The point is not so much to argue for the above cutoff in the string
theory calculation, as to show that these calculations are intrinsically
ambiguous%
\footnote{For related comments see the published version of \citep{Donagi:2008kj}.%
}. 

How does this argument compare with that presented in \citep{Kaplunovsky:1995jw}
which we discussed earlier. As emphasized there the non-universal
and non-harmonic part of the gauge threshold correction must essentially
come from low energy physics, and cannot have any information about
the microscopic details of the UV completion of the theory. Furthermore
according to the discussion in that paper, the string theoretic corrections
in order to match the KL formula, must satisfy \eqref{eq:harmonic}.
The calculation in \citep{Conlon:2009kt} clearly violates this (at
least for the case of branes at orientifold singularities). 

What then is the resolution of this conflict. Firstly the string theory
calculations in \citep{Kaplunovsky:1995jw} are UV finite as they
should be. The expression for the (non-universal) threshold correction
is given by an integral over the fundamental domain $\Gamma$ of the
complex structure of the torus 
\begin{equation}
\Delta_{a}(M,\bar{M})=\int_{\Gamma}\frac{d^{2}\tau}{\tau_{2}}({\cal B}_{a}(\tau,\bar{\tau};M,\bar{M})-b_{a}).\label{eq:DeltaHet}
\end{equation}
 The UV finiteness of the integral is a consequence of the restriction
to $\Gamma$. The integral of the first term is however IR divergent
and the second term is added to cancel this. It should be noted that
the resulting function is independent of any scale and is purely dependent
on the (dimensionless) moduli.

In theories where the gauge sector comes from open strings however
typically there are divergent integrals of the form
\begin{equation}
\Delta_{a}(\frac{\mu^{2}}{\Lambda^{2}};M,\bar{M})=\int_{\Lambda^{-2}}^{\mu^{-2}}\frac{dt}{t}{\cal B}_{a}(t;M,\bar{M})\label{eq:DeltaOpen}
\end{equation}
Now, as in the heterotic case, $\mu\partial_{\mu}\Delta_{a}={\cal B}_{a}(\mu^{-2})\rightarrow b_{a}$
(up to some normalization constant) in the limit of $\mu\rightarrow0$,
since in this limit the sum inside the integral is dominated by massless
states going round the open string loop. While this is undoubtedly
the case in the limit, the question is how far above zero one may
take this infra-red region. Note that unlike in the heterotic case,
here there is also an UV divergence unless one imposes (local and
global) tadpole cancellation. The claim made in \citep{Conlon:2009kt}
is that (writing%
\footnote{Note that in \citep{Conlon:2009kt} what we following \citep{Kaplunovsky:1995jw}
have called $\Delta$ is called $\Lambda_{2}$, and ${\cal B}$ is
called $\Delta$! %
} ${\cal B}_{a}={\cal B}_{a}^{{\cal N}=1}+{\cal B}_{a}^{{\cal N}=2}$),
\begin{eqnarray}
{\cal B}_{a}^{{\cal N}=1} & = & b_{a}^{{\cal N}=1}\Theta(t-M_{s}^{-2}),\label{eq:CP1}\\
{\cal B}_{a}^{{\cal N}=2} & = & b_{a}^{{\cal N}=2}\Theta(t-(RM_{s})^{-2}).\label{eq:CP2}
\end{eqnarray}
In other words the infra-red region is essentially taken (from $\mu^{2}=0$
which is all that is really justified in the string calculation) all
the way to the string scale in the first equation and even beyond
it to the winding scale in the second equation. This is tantamount
to pushing a background field calculation way beyond its regime of
validity. Also the entire modulus dependence of this calculation comes
from the UV cutoff. This is in contrast to the comparisons made between
the KL formula and (UV finite) heterotic orbifold calculations done
in \citep{Kaplunovsky:1995jw} where it is the moduli dependence of
the whole function $\Delta_{a}$ in \eqref{eq:DeltaHet} that is matched
to the KL formula.

In fact of course in type I and its T-dual theories also the corresponding
expressions should be finite. In particular this must mean that redefining
$\Delta_{a}$ as in \citep{Kaplunovsky:1995jw} 
\begin{equation}
\Delta_{a}(M,\bar{M})=\int_{0}^{\infty}\frac{dt}{t}({\cal B}_{a}(t;M,\bar{M})-b_{a}).\label{eq:DeltaOpenMod}
\end{equation}
This is a finite function dependent purely on the moduli once overall
tadpole cancellation is imposed. According to the general arguments
put forth in \citep{Kaplunovsky:1995jw}, it is this function that
should be compared with the one-loop (anomaly) terms of the KL formula
and in particular satisfy \eqref{eq:harmonic}. Any claim to find
a discrepancy must then find a discrepancy with this formula and as
far as we can see this has not been done in the models investigated
in \citep{Conlon:2009kt}. 

Instead what has been done is to split up the divergent integral that
governs tadpole cancellation into two parts as in \eqref{eq:CP1}\eqref{eq:CP2}.
However this integral is finite in the full string theory as a result
of cancellations in the coefficient of the divergent piece - and this
in turn is a consequence of the cancellation of tadpole terms locally
and globally. It is hard to see how a modulus dependence can be extracted
from this cancellation since this coefficient is a pure number independent
of any modulus.

Let us now revisit the issue of linear vs chiral multiplets. In the
geometric regime for the moduli (i.e. with the orbifold/orientifold
singularities blown up) where the field theory makes sense one can
treat the Kähler moduli, either as chiral supermultiplets or as linear
supermultiplets. Unlike the case of the type IIB (heterotic) string
where it naturally arises as a chiral multiplet (linear multiplet)
for the Kähler moduli, one may choose one or the other representation.
In our discussion above in subsection 2.3 we treated all these moduli
on the same footing and got the relation \eqref{eq:TtauGamma} which
we repeat here (after dropping the $S$ dependence for reasons explained
below), 
\begin{equation}
T^{i}+\bar{T}^{i}=\frac{2}{3}\tau^{i}(U)-V^{i}(z,\bar{z}),\label{eq:TtauGamma2}
\end{equation}
Note that with this democratic treatment of all the Kähler moduli,
the arbitrary functions $V^{i}$ cannot depend on the Kähler moduli
(see also \citep{Grimm:2004uq}). Now from the point of view of the
effective field theory of the moduli it is the dynamics that drives
some of the four cycle volumes $\tau^{i}$ to zero (i.e. those for
which $\alpha_{ai}\ne0$). Also if $V^{i}$ are to be identified with
one-loop corrections in string theory then they should be independent
of $S$ as well. Thus this would seem to indicate that any field redefinition
would not involve mixing between different Kähler moduli.

However in \citep{Conlon:2009kt} these moduli are treated asymmetrically.
In other words the claim appears to be that one should only dualize
(to a linear multiplet) only the moduli associated with the shrinking
cycle. Hence these authors get a relation of the form 
\begin{equation}
T^{i}+\bar{T}^{i}=\frac{2}{3}\tau^{i}(U)-V^{i}(z,\bar{z},\tau^{b},\ldots).\,\forall i,\,{\rm with}\,\alpha_{ai}\ne0,\,\alpha_{ab}=0.\label{eq:TtauGammaCP}
\end{equation}
This appears to be highly unnatural from the string derived effective
field theory point of view. One would think therefore that this would
require some strong motivation but as we've discussed above this has
not been firmly established in \citep{Conlon:2009kt}.

\section{Non-perturbative superpotentials and the UV cutoff}

In order to discuss this it is convenient to rewrite the manifestly
supersymmetric superspace supergravity action i.e. \eqref{eq:action},
in a manifestly super-Weyl invariant form;
\begin{eqnarray}
{\cal A} & = & -3\int d^{8}z{\bf E}C\bar{C}\exp[-\frac{1}{3}K(\Phi,\bar{\Phi};{\cal V})]+\nonumber \\
 &  & \left(\int d^{6}z{\cal E}[C^{3}W(\Phi)+\frac{1}{4}f_{a}(\Phi){\cal W}^{a}{\cal W}^{a}]+h.c.\right).\label{eq:WEYLACTION}
\end{eqnarray}
Note that we have also generalized the action slightly in order to
incorporate more than one gauge group and have used the chiral representation
in which $\int d^{2}\bar{\theta}{\bf E}/2R={\cal E}$ the chiral density.
In this form the action has an additional manifest symmetry under
the transformations
\begin{eqnarray}
{\bf E\rightarrow e^{2(\tau+\bar{\tau})}{\bf E},} &  & {\cal E}\rightarrow e^{6\tau}{\cal E}+\ldots,\, C\rightarrow e^{-2\tau}C,\nonumber \\
\nabla_{\alpha}\rightarrow e^{(\tau-2\bar{\tau})}(\nabla_{\alpha}-\ldots), &  & V\rightarrow V,\nonumber \\
\Phi\rightarrow\Phi, &  & {\cal W}_{\alpha}\rightarrow e^{-3\tau}W_{\alpha}.\label{eq:weyl}
\end{eqnarray}
The chiral (auxiliary) field $C$ is introduced so as to make the
Weyl invariance of the theory manifest. It is important to note that
the superpotential occurs in this action with a factor of $C^{3}$.
Now the above (chiral) Weyl transformations are anomalous in the quantum
theory, and the preservation of this local Weyl invariance requires
that the gauge coupling function is changed as follows: \citep{Kaplunovsky:1994fg}
\begin{equation}
f_{a}(\Phi)\rightarrow f_{a}(\Phi)-\frac{3c_{a}}{8\pi^{2}}\ln C,\label{eq:fchange}
\end{equation}
where $c_{a}$ was given in \eqref{eq:ca}. There is however additional
$C$ dependence coming from a field redefinition anomaly which occurs
on demanding canonical normalization for the matter kinetic terms
(see eqn.(16) of \citep{deAlwis:2008aq}). Thus the actual $C$ dependence
changes from the second term of \eqref{eq:fchange} to $-\frac{b_{a}}{8\pi^{2}}\ln C$.
Now suppose that the gauge group becomes strongly coupled and develops
a mass gap below some scale. Then below that scale there is an effective
theory that is obtained by integrating out the gauge theory degrees
of freedom. This gives an effective action $\Gamma$ defined schematically
by%
\footnote{We note that previously we used the letter $\Gamma$ for the effective
guage coupling in the linear multiplet formulation to conform to the
notation in \citep{Binetruy:2000zx}%
} 
\begin{equation}
e^{-\Gamma(\Phi,\bar{\Phi},C,\bar{C})}=\int d(gauge)\exp\left\{ -\frac{1}{4}\int[f_{a}(\Phi)-\frac{b_{a}}{8\pi^{2}}\ln C]{\cal W}^{a}{\cal W}^{a}+h.c.\right\} .\label{GammaEffective}
\end{equation}
Since SUSY should not be broken by this procedure, we expect $\Gamma$
\footnote{The crucial assumption here is the quasi-locality of $\Gamma$ which
enables us to define its derivative expansion and then focus on its
two derivative action which should be be of the standard supergravity
form.%
} to have the general form of a superspace action and in particular
should develop a superpotential. Given the general argument (based
on Weyl invariance above) that any superpotential should come with
a factor $C^{3}$, we see that the corresponding term in $\Gamma$
will be (a superspace integral of) 
\begin{equation}
C^{3}W_{NP}=C^{3}A_{a}\exp\left(-3\frac{8\pi^{2}}{b_{a}}f_{a}(\Phi)\right).\label{eq:WNP}
\end{equation}
Here $A_{a}$ is an $O(1)$ pre-factor (from dimensional analysis
this would mean $A=O(M_{P}^{3})$), which may depend on the moduli
due to threshold corrections. If there is more than one condensing
gauge group, there will obviously be a sum of such terms. This is
essentially the Veneziano-Yankielowicz \citep{Veneziano:1982ah} argument
generalized to SUGRA (see \citep{Burgess:1995aa})%
\footnote{Note that those arguments led to having the anomaly coefficient $c_{a}$
rather than the beta function coefficient $b_{a}$ in $W_{NP}$. Our
discussion shows how $c_{a}$ gets replaced by $b_{a}$.%
}. There is also of course a contribution to the Kaehler potential,
but this is less important since unlike the superpotential, the Kaehler
potential is perturbatively corrected. The total superpotential is
then given as
\begin{equation}
W=W_{c}(\Phi)+W_{NP}=W_{c}(\Phi)+Ae^{-3\frac{8\pi^{2}}{b_{a}}f_{a}(\Phi)},\label{eq:totalW}
\end{equation}
where the first term on the RHS is the classical superpotential. It
should be stressed that this argument is not at all dependent on the
Weyl compensator formalism. If we had worked with $C=1$ (as in Wess
and Bagger \citep{Wess:1992cp}), then the form of the NP term \eqref{eq:WNP}
is what is required to get the right Kaehler transformation of the
superpotential as a result of the Kaehler anomaly (which is now related
to the Weyl anomaly) and renormalization of the matter kinetic term.

On the other hand there is in the literature an alternative form for
$W_{NP}$ based on an RG evolution argument. First one observes that
the IR scale $\Lambda_{a}$ (at which the theory becomes strongly
coupled) is related to the UV scale by 
\begin{equation}
\Lambda_{a}^{3}=\Lambda^{3}e^{-3\frac{2\pi\tau}{b_{a}}}=\Lambda^{3}{\cal V}e^{-3\frac{8\pi^{2}}{b_{a}}\Re f_{a}(\Phi)}.\label{eq:WNPRG}
\end{equation}
Here $\tau\equiv4\pi/g_{a}^{2}$ where $g_{a}$ is the coupling at
the UV scale $\Lambda$. For the first equality we've used the standard
RG argument and for the second equality we've used eqn. \eqref{eq:g_aIIB}
and ignored $O(1)$ corrections to the prefactor.

In the global SUSY literature one often sees the evaluation of the
superpotential as $|W_{NP}|=\Lambda_{a}^{3}$. However in SUGRA as
pointed out in \citep{Conlon:2010ji} this should be replaced by including
a factor which arises from transforming to the Einstein frame. This
comes from the $C^{3}$ factor in \eqref{eq:WNP}, after gauge fixing
$\ln C+\ln\bar{C}=K/3$, the value that is needed to go to Einstein
frame. Thus one should identify
\begin{equation}
e^{K/2}|W_{NP}|=<{\cal W}^{a}{\cal W}^{a}>=\Lambda_{a}^{3}\label{eq:CPed}
\end{equation}
From this after using $K\sim-2\ln{\cal V}$ and \eqref{eq:WNPRG}
we have 
\begin{equation}
|W_{NP}|=\Lambda^{3}{\cal V}^{2}e^{-3\frac{8\pi^{2}}{b_{a}}\Re f_{a}(\Phi)}.\label{eq:Lambdaa}
\end{equation}
Comparing with (the second term of) \eqref{eq:totalW} with $A\sim M_{P}^{3}$
we see that the cutoff $\Lambda$ may be estimated to be
\begin{equation}
\Lambda\sim\frac{M_{P}}{{\cal V}^{2/3}}\sim\frac{M_{string}}{{\cal V}^{1/6}}=M_{KK}\label{eq:LambdaKK}
\end{equation}
In other words the cutoff should be at the KK scale and is safely
below the string scale for large volume. This is in sharp contrast
to the argument of \citep{Conlon:2010ji}.

Finally we point out that if instead of fixing the chiral scalar compensator
$C$ as in the line above eqn \eqref{eq:CPed}, so as to get to Einstein
frame, one imposed $C\bar{C}=\exp(-K/6)\sim{\cal V}^{1/3}$ we actually
get from the SUGRA frame (i.e. with coefficient of $R$ being $e^{-K/3}$)
to the string frame where (at least up to factors of $O(1)$) this
coefficient is $\propto{\cal V}$. Arguably it is in this frame that
the string theoretic calculation should be compared with the SUGRA
expression for the gauge coupling. In this frame the relevant term
of the KL formula (i.e. the second term on the LHS of eqn \eqref{eq:g_aIIB})
becomes 
\begin{equation}
\frac{b_{a}}{16\pi^{2}}\ln\frac{\Lambda^{2}{\cal V}^{1/3}}{\mu^{2}}.\label{eq:stringUnifScale}
\end{equation}
If now we identify the (upper bound on) the apparent unification scale
with the string scale (i.e. $\Lambda{\cal V}^{1/6}\sim M_{string}$
then we again have \eqref{eq:LambdaKK}, i.e. the cutoff should be
the Kaluza-Klein scale.

\section{Conclusions}

In this note we rederived the relation between the chiral and linear
multiplets in the standard supergravity frame. This is not only simpler
than that given in \citep{Binetruy:2000zx} but also clarifies the
relation to the anomaly terms in the KL formula. In the heterotic
string, there is a universal one-loop correction coming from the string
calculation, which can be reinterpreted in the chiral superfield formulation
as a correction to the Kaehler potential. In the case of non-universal
corrections to the shrinking cycle of LVS constructions, we have argued
that the necessity for such redefinition has not been clearly established.
We have also shown, using two different arguments for the non-perturbative
superpotential, that the effective cutoff is around the KK scale,
in agreement with low energy effective action expectations. This resolves
the apparent disagreement between the two different expressions that
have appeared in the literature for these NP terms.

\section{Acknowledgments}

I wish to thank Fernando Quevedo and Joe Conlon for discussions, and
the Newton Institute Cambridge and the organizers of the BSM workshop,
for providing an excellent atmosphere for discussing the issues addressed
in this paper. The award of an SFB fellowship from the University
of Hamburg and DESY, and a visiting professorship at the Abdus Salam
ICTP are also gratefully acknowledged. This research is partially
supported by the United States Department of Energy under grant DE-FG02-91-ER-40672. 

\[
\]
 \bibliographystyle{apsrev} \bibliographystyle{apsrev}
\bibliography{myrefs}

\end{document}